\documentclass[preprint]{vis/vgtc}             

\ifpdf
  \pdfoutput=1\relax                   
  \usepackage{graphicx}                
  \DeclareGraphicsExtensions{.pdf,.png,.jpg,.jpeg} 
\else
  \usepackage{graphicx}                
  \DeclareGraphicsExtensions{.eps}     
\fi%

\graphicspath{{figures/}{pictures/}{images/}{./}} 

\usepackage{microtype}                 
\PassOptionsToPackage{warn}{textcomp}  
\usepackage{textcomp}                  
\usepackage{mathptmx}                  
\usepackage{times}                     
\usepackage{cite}                      
\usepackage{tabu}                      
\usepackage{booktabs}                  

\onlineid{1168}

\vgtccategory{Research}
\vgtcpapertype{evaluation}

\vgtcinsertpkg

\title{Automatic Y-axis Rescaling in Dynamic Visualizations}

\author{Jacob Fisher\thanks{e-mail: jacob.fisher@columbia.edu}\\ %
        \scriptsize Columbia University %
\and Remco Chang\thanks{e-mail: remco@cs.tufts.edu}\\ %
     \scriptsize Tufts University %
\and Eugene Wu\thanks{e-mail: ewu@cs.columbia.edu}\\
        \scriptsize Columbia University %
        }

\teaser{
  \centering
  \includegraphics[width=\linewidth]{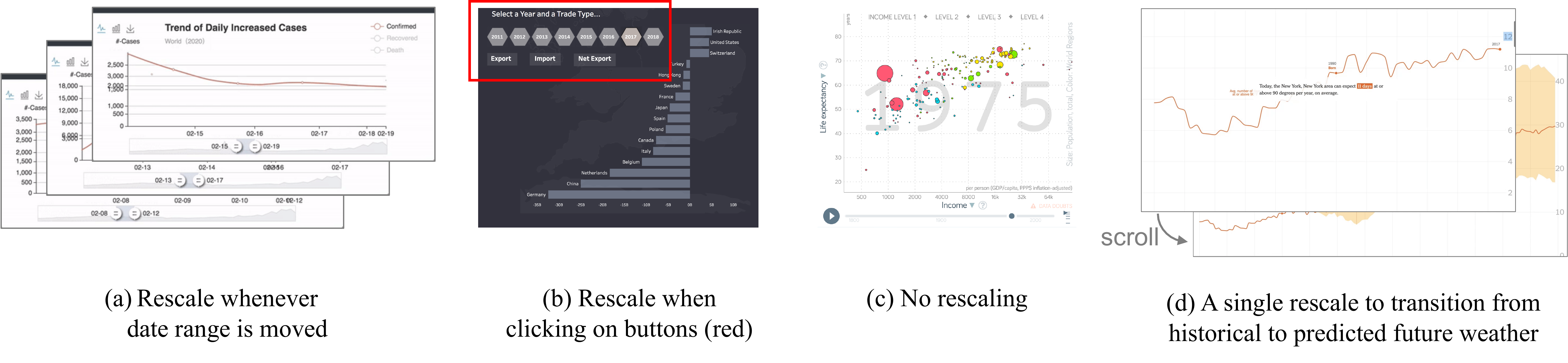}
  \caption{Four interactive visualizations that have different rescaling policies for the y-axis.  (a) COVID-19 infection statistics rescales whenever the chart data changes in response to range slider interactions. (b) UK Balance of Trade on Tableau Public rescales whenever the user clicks a different year or statistic.  Note chart is flipped, so the ``y-axis'' is horizontal.  (c) Gapminder visualization uses fixed axes that are not rescaled during the animation and slider interactions.  (d) New York Times animates in temperature over time as the user scrolls down, and rescales the y-axis to transition from historical to future temperatures.}
  \label{fig:teaser}
}

\abstract{
Animated and interactive data visualizations dynamically change the data rendered in a visualization (e.g., bar chart).  As the data changes, the y-axis may need to be rescaled as the domain of the data changes.  Each axis rescaling potentially improves the readability of the current chart, but may also disorient the user. In contrast to static visualizations, where there is considerable literature to help choose the appropriate y-axis scale, there is a lack of guidance about how and when rescaling should be used in dynamic visualizations.  Existing visualization systems and libraries adapt a fixed global y-axis, or rescale every time the data changes.   Yet, professional visualizations, such as in data journalism, do not adopt either strategy.  They instead carefully and manually choose when to rescale based on the analysis task and data.
To this end, we conduct a series of Mechanical Turk experiments to study the potential of dynamic axis rescaling and the factors that affect its effectiveness. We find that the appropriate rescaling policy is both task- and data-dependent, and we do not find one clear policy choice for all situations.
}

\CCScatlist{
  \CCScatTwelve{Human-centered computing}{Visualization}{Empirical studies in visualization}{};
  \CCScatTwelve{Human-centered computing}{Visualization}{Visualization theory, concepts and paradigms}{}
}

\usepackage{amsmath}
\usepackage[a-1b]{pdfx}
\usepackage{amsfonts}
\usepackage{amssymb}
\usepackage{multirow}
\usepackage{enumitem}
\usepackage{listings}
\usepackage{booktabs} 
\usepackage{xspace}
\usepackage{alltt}
\usepackage{xcolor}
\usepackage{microtype}
\usepackage{mathtools}
\usepackage{subcaption}
\usepackage{balance}
\usepackage{tikz}
\usepackage[noend]{algpseudocode}
\usepackage{algorithmicx,algorithm}
\usepackage{comment}
\usepackage{microtype}
\usepackage{pgffor}
\usepackage{mwe}
\usepackage{todo}

\newtheorem{example}{Example}

\newlength{\listingindent}                
\usepackage[LGRgreek]{mathastext}

\usepackage[noabbrev,capitalise]{cleveref}

\definecolor{light-gray}{gray}{0.95}
\definecolor{mid-gray}{gray}{0.85}
\definecolor{darkred}{rgb}{0.7,0.25,0.25}
\definecolor{darkgreen}{rgb}{0.15,0.55,0.15}
\definecolor{darkblue}{rgb}{0.1,0.1,0.5}
\definecolor{blue}{rgb}{0.19,0.58,1}

\lstdefinestyle{SQLStyle}{
  language=SQL,
  showspaces=false,
  basicstyle=\ttfamily\scriptsize,
  commentstyle=\color{gray},
  mathescape=true,
  numbers=none,
  escapeinside={^}{^},
  captionpos=b,
  float=tp,
  floatplacement=tbp,
  belowskip=-0.05em,
   mathescape=false
}

\newcommand{\eat}[1]{}

\newcommand{\stitle}[1]{\vspace{2pt}\noindent\textbf{#1}}

\begin{document}

\firstsection{Introduction}

\maketitle

Scales and axes provide powerful contextual cues to help users interpret data visualizations.    Considerable prior work has established guidelines to choose the appropriate scale transformation~\cite{wilkinson2012grammar,Cleveland1985TheEO} (e.g., linear, log),  set the domain of the scale based on the data~\cite{wang2017there}, choose ticks and labels~\cite{Talbot2010AnEO,wilkinson2012grammar,stirling1981algorithm}, and maintain consistency of scales between multiple views~\cite{Qu2018KeepingMV}. In addition, automatic visualization tools can help recommend the appropriate scale transformations~\cite{Mackinlay2007ShowMA,Moritz2019FormalizingVD,Mackinlay1986AutomatingTD}. These studies were primarily in the context of static visualizations.

In contrast, dynamic visualizations---such as animated or interactive visualizations---change the rendered data as the animation progresses, or in response to user interactions.  Each frame of the visualization presents a different set of data in the chart.   A naive approach is to use existing guidelines to use a separate y-axis scale for each frame.  (This paper uses the term ``y-axis'' to refer to the encoding channel for the visualization's measure attribute.)  However, a y-axis that changes every frame removes a shared reference frame and can disorient the user.  At the other extreme, the visualization may use a single global y-axis whose domain contains the minimum and maximum of all possible data that can be rendered in the visualization.  However, this can obscure the data in any given frame. The following is an example of this trade-off:

\begin{example}[Interactive COVID-19 Visualization]\textit{
  \Cref{fig:teaser}(a) is an interactive visualization that depicts the number of COVID-19 infections over a 4 day window.  The window can be moved by dragging the date range an the bottom of the visualization.  The figure depicts three adjacent windows in February when the number of cases spikes from 3500 (bottom frame) to 18000 to 2500 (top frame).  In this visualization, the y-axis rescales whenever the data changes (when the slider moves).  This always ensures that the data uses the full height of the chart, but can disorient the user.  However, a fixed scale of $[0, 18000]$ makes it difficult to perceive the changes in data in the bottom and top frames.}
\end{example}

Qu et al.~\cite{Qu2018KeepingMV} proposed a consistency constraint model to evaluate how authors maintain encoding and scale consistency across multiple static views.  They propose a consistency criteria that the same data field should be encoded using the same scales (the same data domain and retinal range).      If directly applied to dynamic visualizations, where the retinal range and the data fields remain constant, this criteria implies that the domain of the scales should be fixed across the frames.  On the other hand, their participants tended to ignore this consistency criteria when it would lead to ``\textit{too much whitespace}'' that would result in  ``\textit{leaving details and trends harder to see in that view}''.   This suggests situations where rescaling may be desirable.

Choosing the appropriate rescaling strategy is challenging because there are many possible rescaling policies.  For a chart parameterized by a slider with $n$ slider values, there are $2^{n-1}$ possible rescaling policies: each frame be rescaled (which we call a {\it breakpoint}) or reuse the previous frame's scale. 

In practice, existing visualization systems such as Tableau~\cite{tableau}, and libraries such as Vega-lite~\cite{Satyanarayan2017VegaLiteAG}, use one of two extreme policies: rescale on every change, or never rescale by using a fixed axis.  However, these policies may not be optimal.  Professionally designed interactive visualizations often use intermediate policies.  For instance, \Cref{fig:teaser}(d) is a New York Times climate article that depicts temperature from 1960 to (projected) 2090.  As the user scrolls down, data is animated in from 1960 to 2017; the y-axis is then rescaled to signify the contrast between historical and predicted future temperatures.

In this paper, we conduct a preliminary crowdsourced study on rescaling policies.  
Participants perform visual analysis tasks by interacting with a slider widget  (\Cref{fig:pilotV1UI}).   Our motivation is to develop rescaling guidelines for designers that are building dynamic visualizations or for automatic interactive visualization design systems~\cite{tableau,Satyanarayan2017VegaLiteAG,zhang2019mining}.  

Our main hypothesis is that the effects of a given rescaling policy depends on the type of analysis task (H1).  Specifically, rescaling on every frame may improve the accuracy of tasks that compare data within individual frames, whereas using a fixed global scale is beneficial for tasks that compare across frames.   Our second hypothesis is that an ``intermediate'' policy---where it only rescales when the frame's y-axis data domain changes considerably---is beneficial across different analysis tasks (H2).

Regarding H1, our studies find that axis rescaling affects participant task accuracy and/or latency depending on the dataset and task, at least in the line chart examples examined.  For datasets that do {\it not} exhibit large variations, global rescaling is generally effective across tasks, while per-chart rescaling negatively affects latency and accuracy.  However, when the variation is larger, per-chart rescaling can benefit tasks that require comparing marks within a chart.  
Regarding H2, breakpoint policies appear to strike a robust middle-ground across data and tasks: although they do not clearly out-perform global nor per-chart, they also do not clearly perform worse.  These findings suggest that axis rescaling is a promising tool when designing interactive visualizations, and that automatic methods for choosing breakpoints can be helpful.

\section{Studies Overview}
We conducted a series of three Mechanical Turk user studies to compare different rescaling policies in an interactive slider-based visualization: a single global scale, per-chart scaling method, and breakpoint policies that vary the number of slider positions where the y-axis is rescaled (a breakpoint).
All three studies were designed to test both H1 and H2.  After each study, we re-assessed participants' reactions to our data, tasks, and conditions, and altered them accordingly for later studies.  Specifically, we modify our data to amplify extremities and remove a breakpoint condition for the second study, as it did not provide new information from another condition. For the third study, we again modify our data, removed a task, and replaced one breakpoint condition with a breakpoint condition that includes markers for breakpoints, in order to test if these markers make breakpoints easier to use.  Additionally, we run the third study as within-subjects.

\begin{figure}[htb]
  \centering
  \includegraphics[width=.8\linewidth]{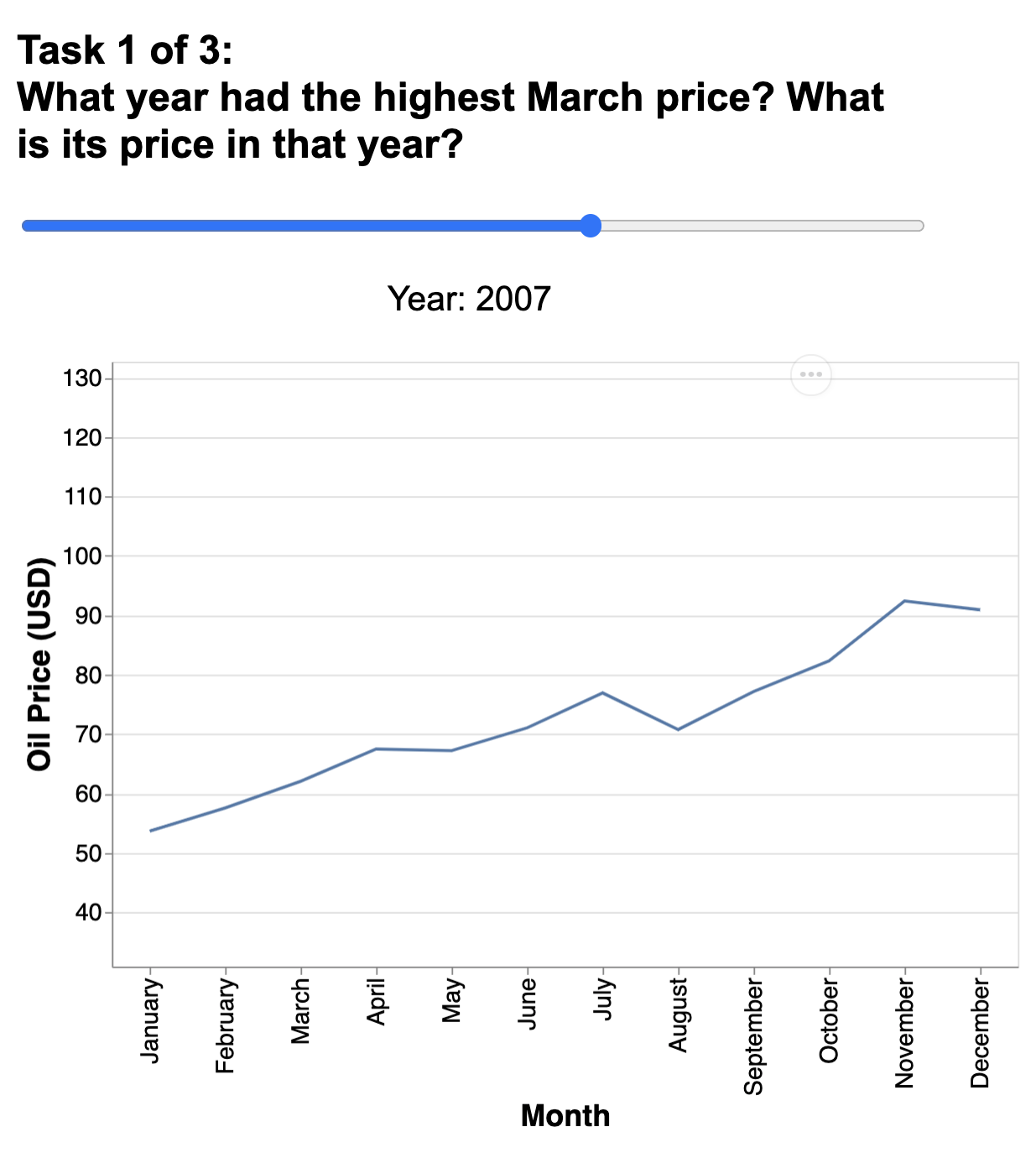}
  \caption{\label{fig:pilotV1UI}
           The user interface for the pilot study.}
\end{figure}

\stitle{Shared Protocols}
In all studies, participants were presented  with a line chart and a slider that controls the chart data (\Cref{fig:pilotV1UI}).  
The chart displays monthly crude oil prices for a given year, and the slider varies the year.  We manually selected the breakpoints so that the variation in the data along the y-axis was minimized for the frames between the breakpoints.
To familiarize participants with the interface, they first complete 4 qualification tasks where they are asked to manipulate the slider and state whether the y-axis scale changes at $1-5$ times, ${>}5$ times, or does not change.  
In each task, the participant clicks a start button to load the visualization and start the timer.  She then clicks a ``ready to submit'' to hide the chart and enter the answers to the task.  We further ask for qualitative feedback and comments on the task itself.  At the end of the study, participants are asked a brief demographic survey (age group and education level).
We paid participants equivalent to \$15/hr for finishing all tasks.  This came out to \$1.50 for studies 1 and 2 and \$4 for study 3.

\subsection{Pilot Study}

We first ran a between-subjects pilot study to understand the extent different policies affect a participant's task performance, and to assess three types of tasks.  We used 5 policies: global, per-chart, and either 1, 3, or 5 breakpoints.  This study was run as a between-subject study.  Participants were randomly assigned to one policy.
Participants were first given a training/qualification task to familiarize them with the interface and their assigned rescaling policy.  
The task asked ``Is there a month in 1999 where the oil price was above \$75?'' and is unrelated to the main tasks.  

Participants then completed three tasks in random order.
(T1) ``What year had the highest March price?  What is its price in that year?''; 
(T2) ``What is the difference in prices between April 1998 and September 1998?'';
(T3)``In how many years was the price for August within \$5 of the price for February?''
Task 1 compares data across frames, and we hypothesized the global policy would be best (H1.1). 
Task 2 compares within a frame, and we hypothesized per-chart would be best (H1.2).
Task 3 compares both within and between frames, and we hypothesized the breakpoint policies would be best (H1.3), though we were not certain which variant.  
We used the Brent Oil Prices dataset from 1988 to 2018 because of large oil price fluctuations in the dataset.  In particular, the oil prices start relatively low, rise by an order of magnitude, and then fluctuate quite a bit.  Our motivation for using this type of dataset was that the global policy could perform poorly because prices in earlier years would appear ``squashed''.

\begin{figure*}
  \centering\hfill
  \begin{subfigure}[b]{0.32\textwidth}
    \centering
    \includegraphics[width=\linewidth]{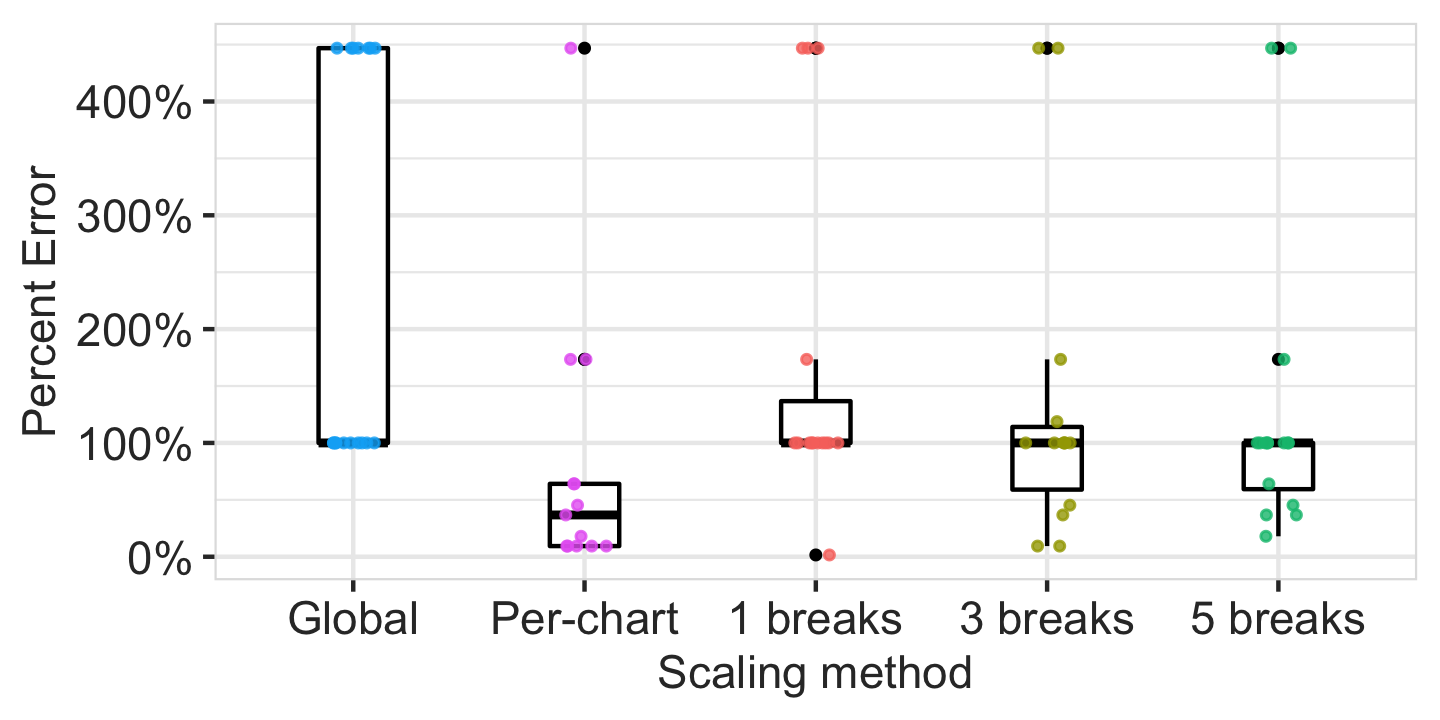}
    \caption{\label{fig:error_t2_study1}
    Response accuracy, Task 2.}
  \end{subfigure}\hfill
  \begin{subfigure}[b]{0.32\textwidth}
    \centering
    \includegraphics[width=\linewidth]{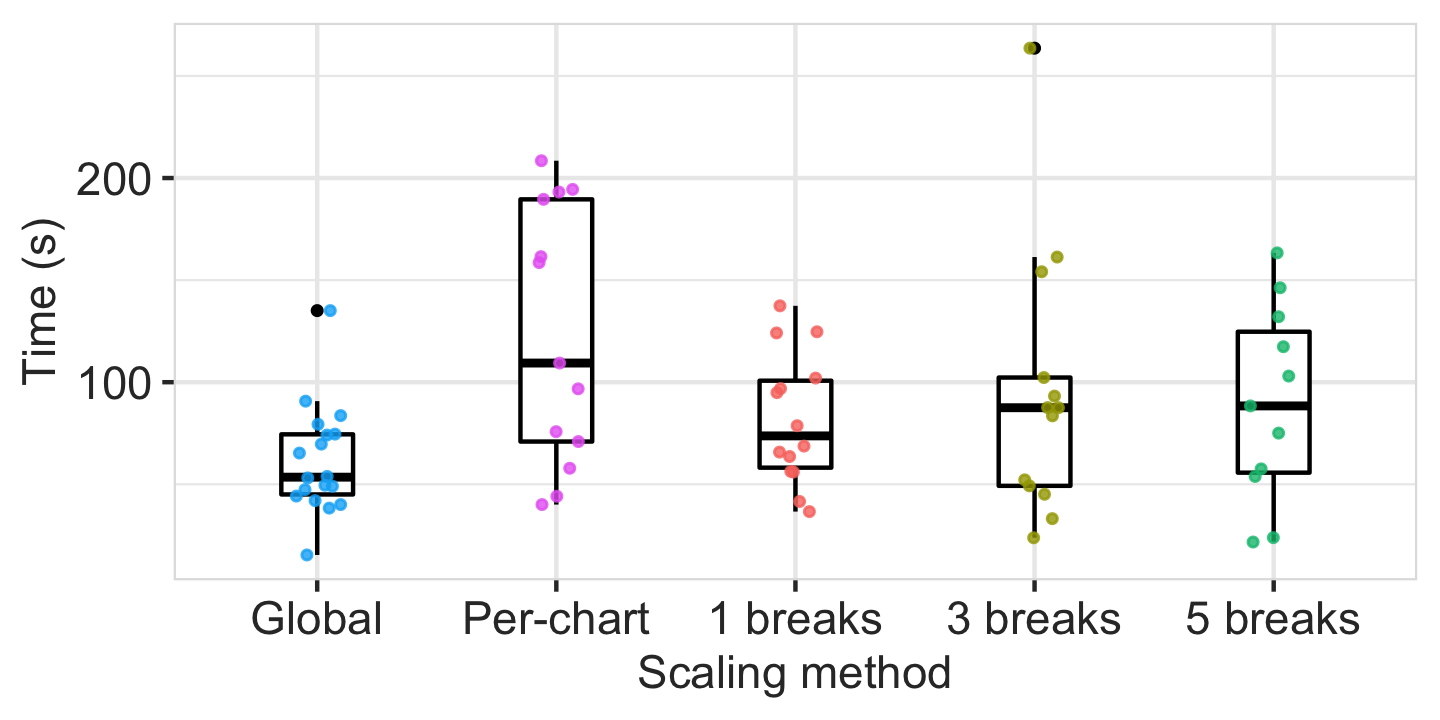}
    \caption{\label{fig:time_t1_study1}
    Response latency, Task 1.}
  \end{subfigure}\hfill
  \begin{subfigure}[b]{0.32\textwidth}
    \centering
    \includegraphics[width=\linewidth]{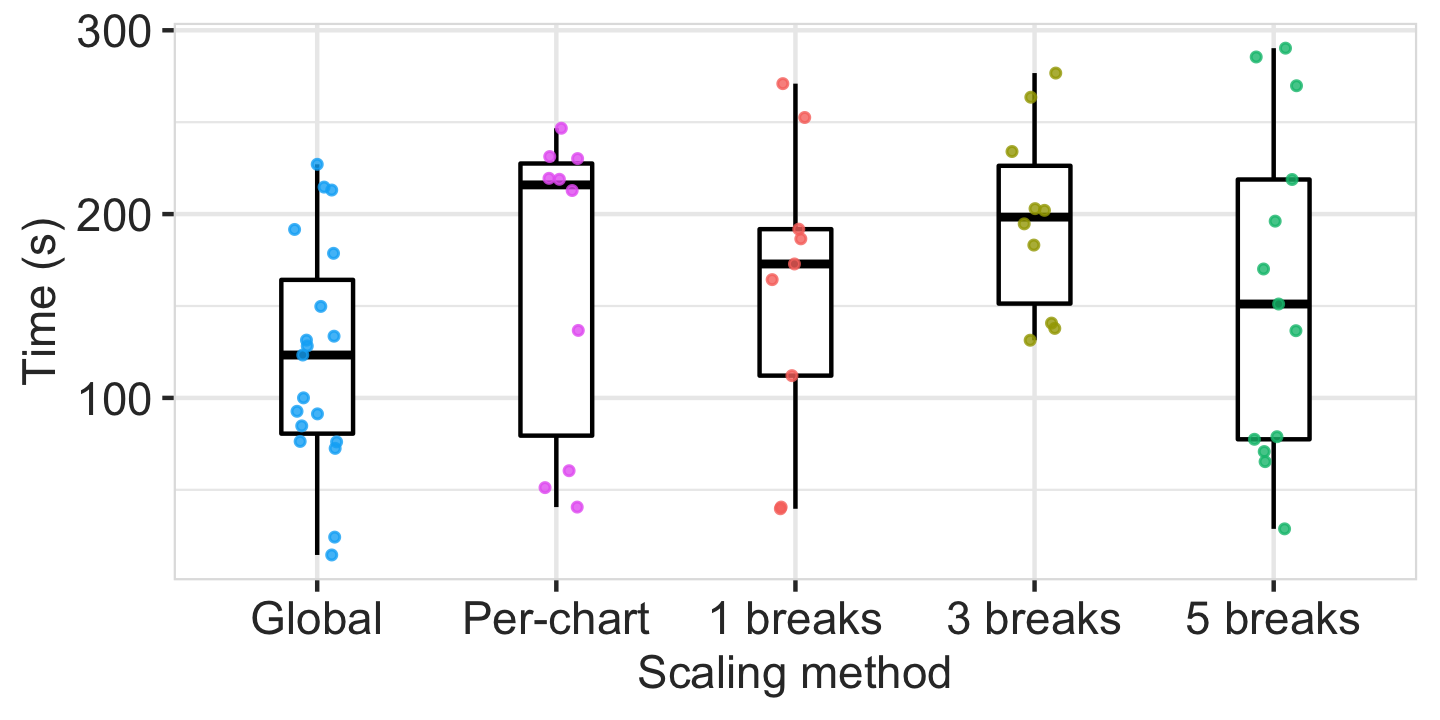}
    \caption{\label{fig:time_t3_study1}
    Response latency, Task 3.}
  \end{subfigure}
  \hfill
  \caption{Results from first Pilot study.}
\end{figure*}

\subsubsection{Results}
We collected data from 88 participants.  We removed the top and bottom 10\% of task answers, and responses completed within 5 seconds.
We report one-way ANOVA tests for latency and relative response error, where the factor is the scaling policy.  $p$-values under the 0.05 level of significance are in bold.
Note that Task 1 asks participants to estimate the price and year, though the middle 80\% of responses for year were all correct, so ANOVA was not run for that data point.

\begin{table}[tb]
  \scriptsize%
	\centering%
  \begin{tabu}{r r l l l}
    \toprule
    Study & Task & Latency & Error (price) & Error (year) \\
    \midrule
    1 & 1 & \bf{0.017} & 0.626 & N/A \\
    1 & 2 & 0.831 & \bf{0.045} & - \\
    1 & 3 & \bf{0.016} & 0.868 & - \\
    2 & 1 & \bf{0.025} & 0.405 & N/A \\
    2 & 2 & 0.702 & \bf{0.000} & - \\
    2 & 3 & 0.337 & 0.768& - 

  \end{tabu}
  \caption{P-values from ANOVA in first and second pilot studies.}
  \label{tab:study12pval}
\end{table}

\Cref{tab:study12pval} shows that the rescaling policy was not significant for latency nor error in Task 1, due to the simplicity of the task.
We found that latency, error, and latency were significant for tasks 1, 2, and 3 respectively.
\Cref{fig:error_t2_study1} shows, as expected from H1.2, that error under global performs worse due to the large variation in the data across the years, whereas the breakpoint conditions appear slightly worse than per-chart.
\Cref{fig:time_t1_study1} and \Cref{fig:time_t3_study1} show that for these tasks,  per-chart indeed slows participants down, and breakpoints are in between global and per-chart.

Overall, we found that the rescaling policy can affect task completion time and accuracy, and it depends on the task type.
However, the range of the oil prices is only between \$9.82 to \$132.7, and we decided to evaluate data with larger variations to more clearly assess the differences between the policies (if any).

\subsection{Second Pilot Study}

We followed the procedure of the Pilot study, but removed the 1 breakpoint policy as it appeared identical to the 3 breakpoint policy.  We again ran this study as a between-subjects study.  
We modified the dataset to greatly accentuate oil pricing differences---for each price $p$, we replaced its value with $1.15^{p}$.  
Based on this data, we updated Task 2 to ``What is the difference in prices between April 2000 and September 2000?'', and Task 3 to 
``In how many years was the price for August within \$10 of the price for February?''.

\subsubsection{Results}
\begin{figure}[htb]
  \centering
  \includegraphics[width=.8\linewidth]{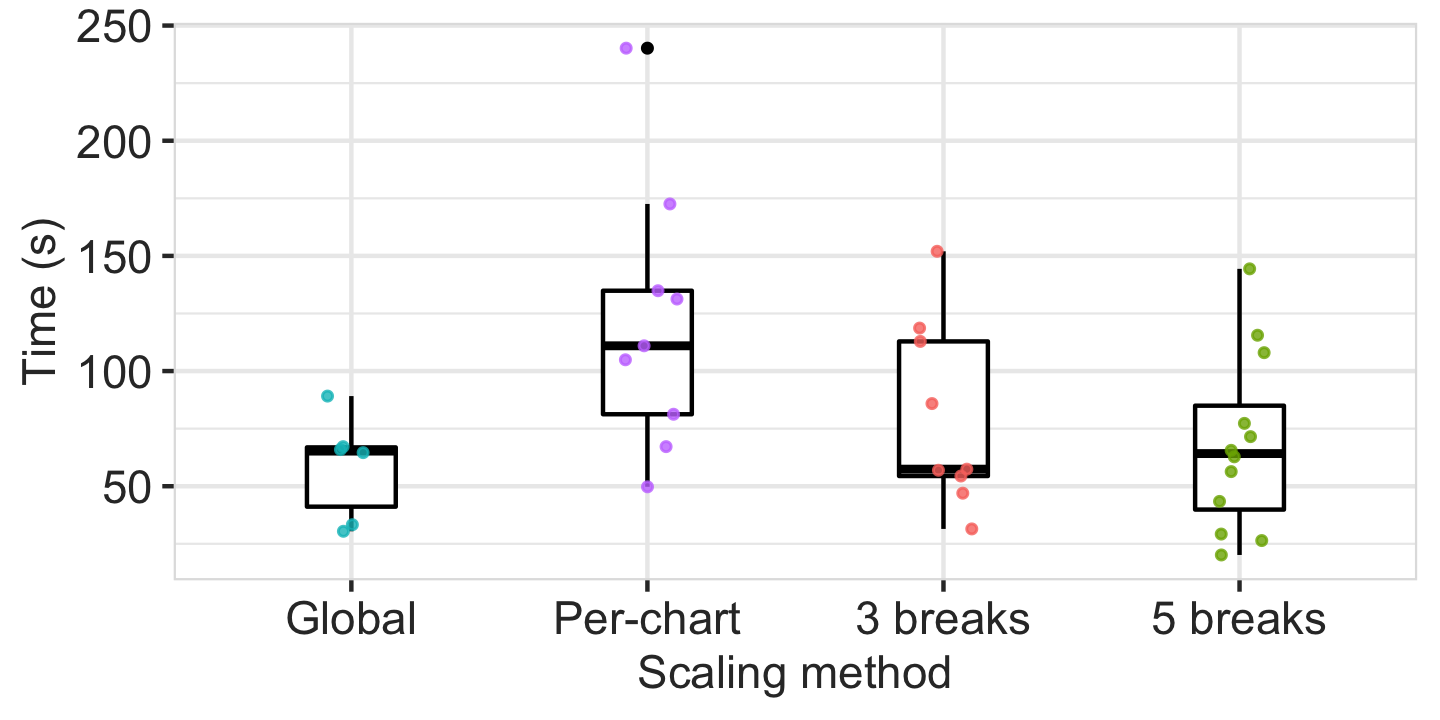}
  \caption{\label{fig:time_t1_study2}
  Response latency, Task 1, Second Pilot.}
\end{figure}

\begin{figure}[htb]
  \centering
  \includegraphics[width=.8\linewidth]{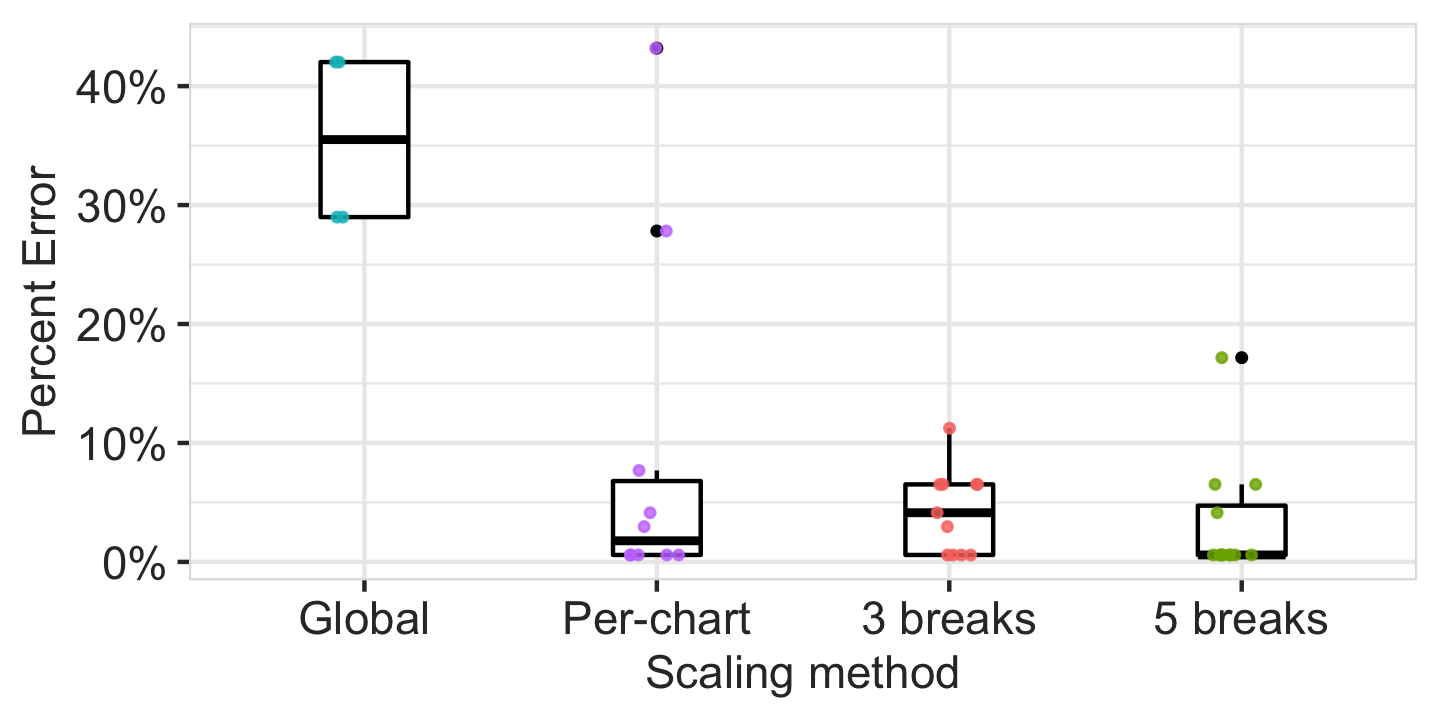}
  \caption{\label{fig:error_t2_study2}
  Response accuracy, Task 2, Second Pilot.}
\end{figure}

We collected data from 46 participants and ran one-way ANOVA on the 10\% trimmed means for each task.  
\Cref{tab:study12pval} shows that the rescaling policy has a significant effect on latency in Task 1, but not accuracy.  
\Cref{fig:time_t1_study2} shows, as expected, that global is fastest, followed by 3 and 5 breakpoints, and finally per-chart.  
Per-chart is slowest because the participant must repeatedly assess the y-axis because it changes every frame.  
We again found a significant effect for error on Task 2 (Figure~\ref{fig:error_t2_study2}), and the difference between global and the other policies is more pronounced, as expected due to the data amplification.  
We were surprised that the breakpoint policies had no effect despite designing Task 3 to favor them.  
Participant comments consistently stated Task 3 as too difficult (``I found it difficult to keep count of just how many years fit the criteria.''),
and so we removed it in the final study.  
Participants also commented that breakpoints were challenging because the rescaling was  unexpected when scrubbing the slider, and could disrupt their analysis:  (``The scale change was sometimes hard to notice. It is not something you would expect to change so I feel it is easy to miss.'')
Thus, we added design considerations in the final study.

\subsection{Third Study, With Markers}

\begin{figure*}
  \centering\hfill
  \begin{subfigure}[b]{0.32\textwidth}
    \centering
    \includegraphics[width=\linewidth]{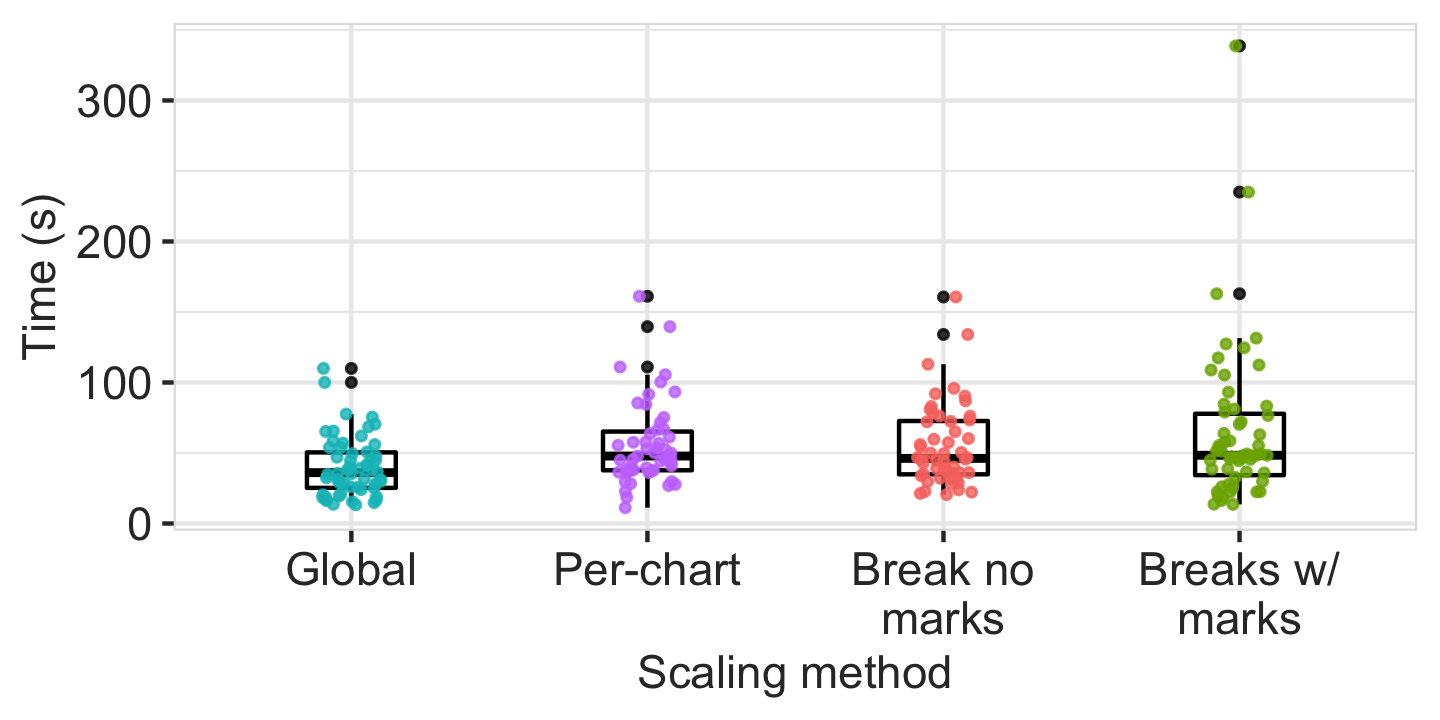}
    \caption{\label{fig:time_t1_study3}
    Response latency, Task 1.}
  \end{subfigure}\hfill
  \begin{subfigure}[b]{0.32\textwidth}
    \centering
    \includegraphics[width=\linewidth]{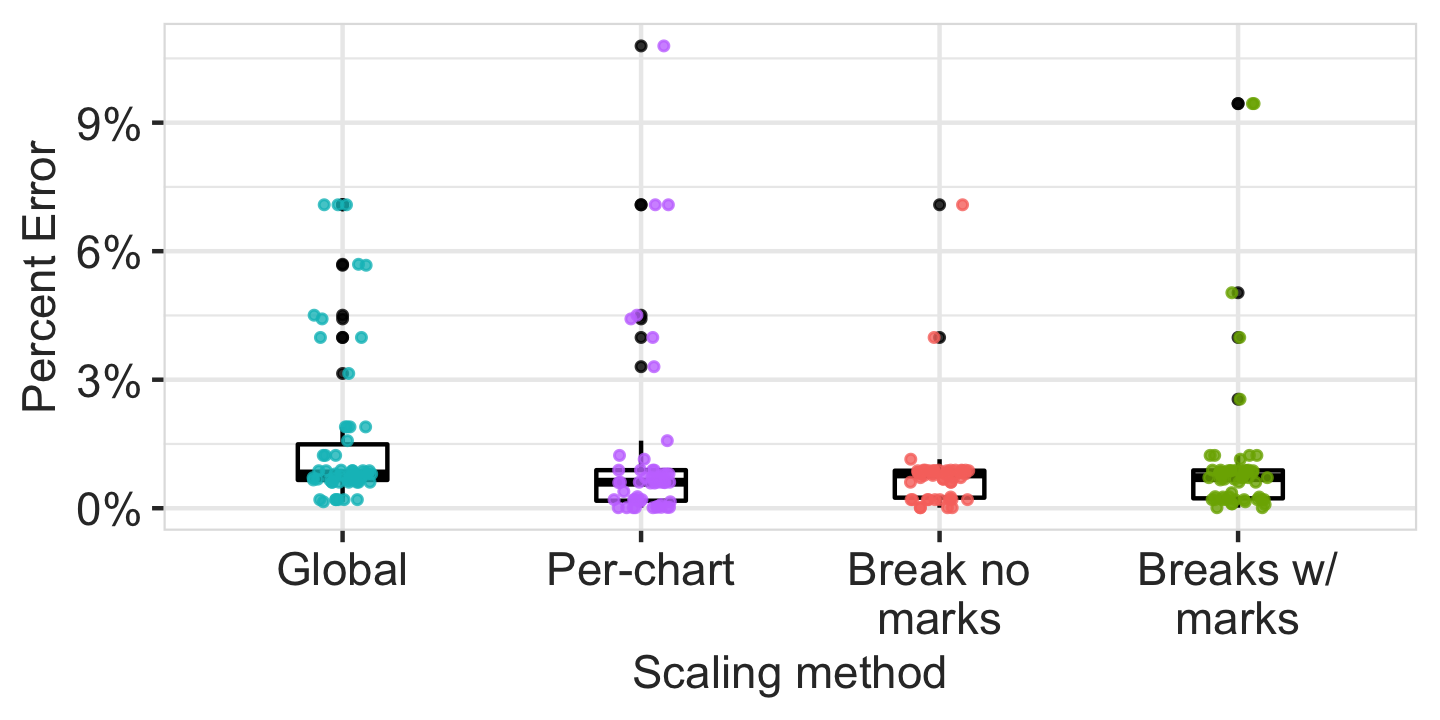}
    \caption{\label{fig:error_t1_study3}
    Response accuracy, Task 1.}
  \end{subfigure}
  \begin{subfigure}[b]{0.32\textwidth}
    \centering
    \includegraphics[width=\linewidth]{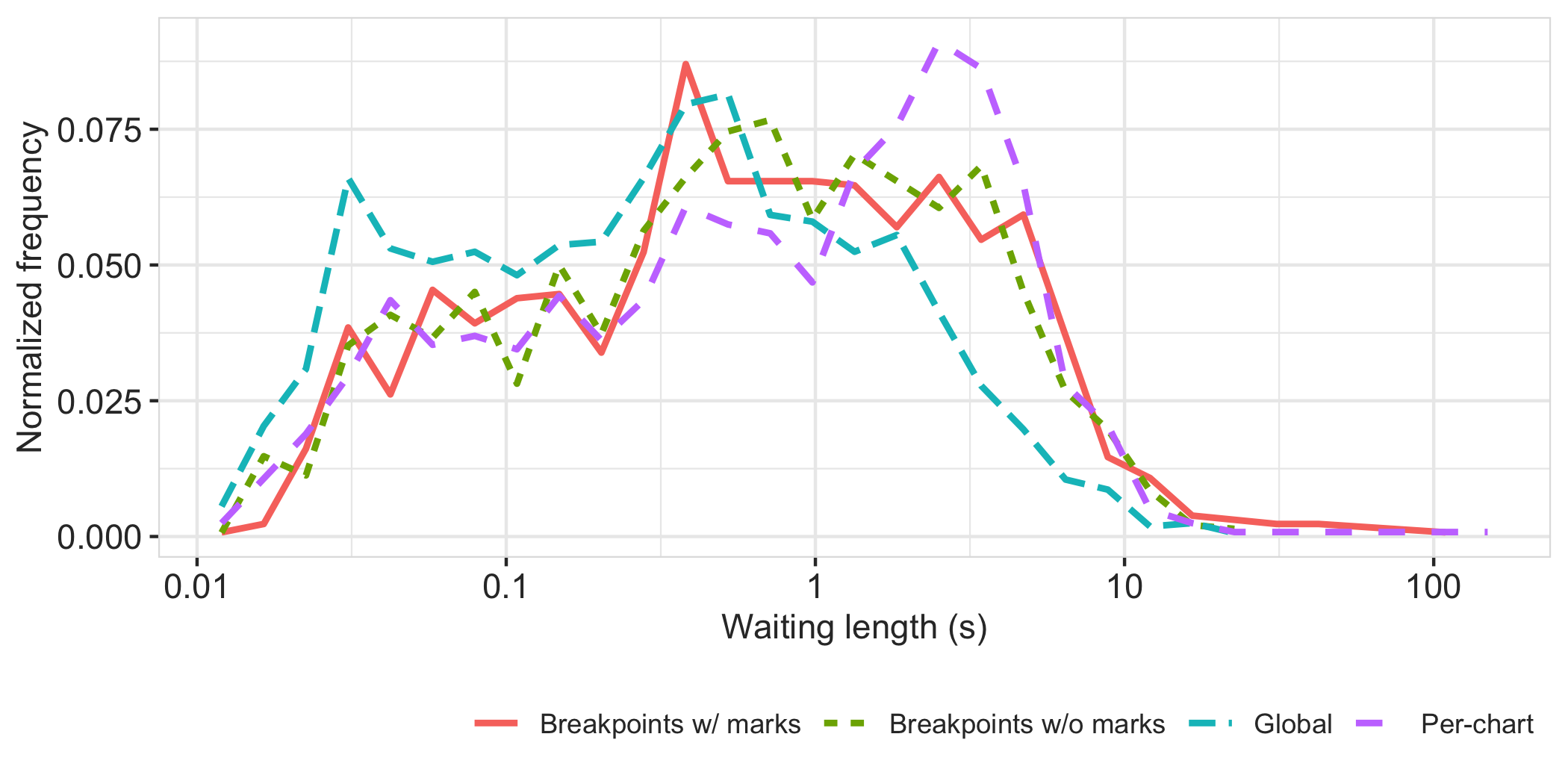}
    \caption{\label{fig:histogram}
    Distribution of waiting lengths, Task 1.}
  \end{subfigure}
  \caption{Results from third study}
\end{figure*}

\begin{figure}[htb]
  \centering
  \includegraphics[width=.8\linewidth]{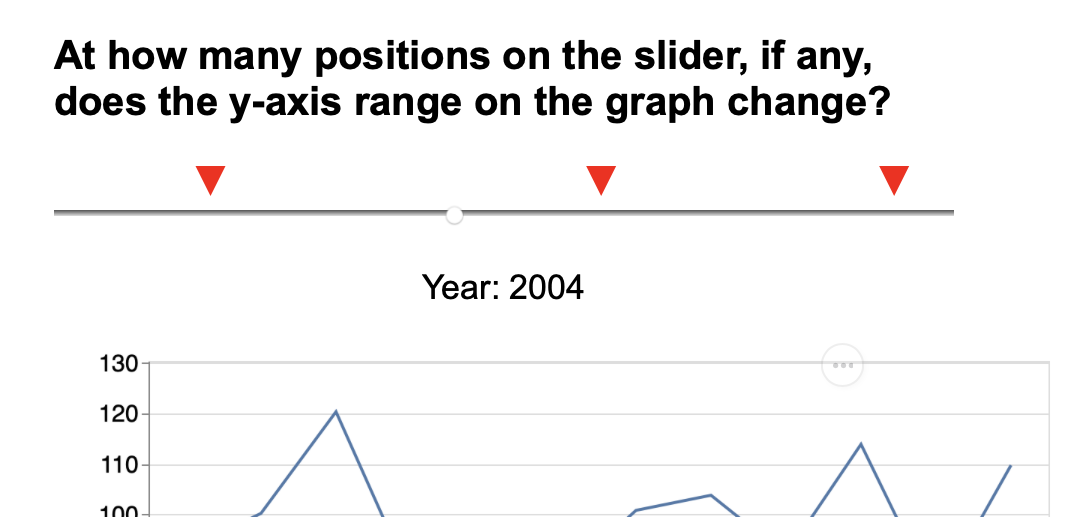}
  \caption{\label{fig:study3}
  Visual indicators on the slider denoting the locations of breakpoints.}
\end{figure}

Based on the prior studies, we ran a final study, now using a within-subjects design.  Participants completed 8 tasks in total: they completed Tasks 1 and 2 (we removed Task 3) using the following policies: global, per-chart, 3 breakpoints without markers, and 3 breakpoints with markers, all in a randomized order.
\Cref{fig:study3} shows the markers placed at each breakpoint location between the two adjacent frames.
The specific task questions were:
(T1) ``What was the highest price for October?'' and (T2) ``What is the difference in prices between May 2005 and June 2005?''.
Note that we only ask for the highest price in October, rather than the price and year.  

Since participants complete the same tasks under multiple policies, using the same data would be subject to learning effects.  Thus, we generated synthetic data for 10 years (2000 to 2009) that had similar characteristics as the oil dataset, but where the specific fluctuations were randomly generated.
We also updated the qualification tasks so participants were familiarized with each of the scaling methods and asked to answer ``At how many positions on the slider, if any, does the y-axis range on the graph change?''
To better understand how the policies affected how participants interact with the visualization, we logged interaction traces by saving the timestamp and location of every change to the slider.
Finally, we asked participants to simply rank the policies from 1 (easiest) to 4 (hardest to use).

\subsubsection{Results}

\begin{table}[tb]
  \scriptsize%
	\centering%
  \begin{tabu}{r l l}
    \toprule
    Task & Latency & Error \\
    \midrule
    1 & \bf{0.034} & \bf{0.001} \\
    2 & 0.168 & 0.530 
  \end{tabu}
  \caption{Results of ANOVA tests in the third study.}
  \label{tab:study3pval}
 \end{table}

\begin{table}[tb]
  \scriptsize%
	\centering%
  \begin{tabu}{l l l}
    \toprule
    Rescaling Policy 1 & Rescaling Policy 2 & $p$-value\\
    \midrule
    Breakpoints w/ marks & Breakpoints no marks & 0.397\\
    Breakpoints w/ marks & Global & \bf{0.000}\\
    Breakpoints w/ marks & Per-chart & \bf{0.002}\\
    Breakpoints no marks & Global & \bf{0.000}\\
    Breakpoints no marks & Per-chart & \bf{0.000}\\
    Global & Per-chart & \bf{0.000}\\
  \end{tabu}
  \caption{Results of Kolmogorov–Smirnov test comparing distribution of interaction wait lengths in third study Task 1.}
  \label{tab:histogrampval}
 \end{table}
We collected data from 66 participants, and report one-way within-subjects ANOVA on the 10\% trimmed means (\Cref{tab:study3pval}). 
Both latency (\Cref{fig:time_t1_study3}) and error (\Cref{fig:error_t1_study3})  were significant in Task 1.  
Although latency was consistent with the pilots, error was unexpected, as per-chart performed significantly worse than the alternatives.  
We conjecture that an explanation could be that per-chart requires more effort, participants may have chosen an incorrect year and thus made a mistake.
Finally, although we added visual markers for the breakpoints based on participant feedback in the pilots, we did not find a significant effect in performance, however participants ranked visual markers higher on average (2.24) than without markers (3.03).

We also studied whether the rescaling policies affect {\it how} participants interact with the visualization in Task 1.  Specifically, the breakpoint methods would stabilize the y-axes in between the breakpoints, and thus may invite faster scrubbing for analyses like Task 1.
To do so, we measure the time between interaction events, and compute the distribution of these wait times (\Cref{fig:histogram}).    
If participants scrubbed more quickly when the axis scale doesn't change, then we would expect long waiting lengths to be less frequent under the global and breakpoint policies.  
We in fact see that global has a higher frequency of very short wait lengths, while participants with per-chart spent the greatest portion of their analysis waiting between $1-10$ seconds.  Breakpoints were between the two conditions.  
We ran pair-wise Kolmogorov–Smirnov tests between the four distributions, and used Bonferroni correction with a factor of 6 (significance threshold is 0.0083).  Table~\ref{tab:histogrampval} shows that all pairs were significant, aside from the two breakpoint policies.  
\section{Conclusion and Future Directions}

This paper conducted an evaluation of dynamic y-axis rescaling strategies for interactive line chart visualizations, where the user sees a sequence of ``frames'' in response to user interactions.  In addition to commonly used existing strategies that use a fixed y-axis (global) or rescale the y-axis whenever the chart data updates (per-chart), we study a {\it breakpoints} strategy that rescales the y-axis based on the magnitude that the data changes between successive frames.  We find that global and per-chart are respectively well-suited for tasks that compare data across frames and within a single frame, however they perform poorly for the other tasks and depend on how the data's scale changes.  Further, {\it breakpoints} is a robust strategy that performs comparably to the best strategy across tasks.

A promising future direction is an algorithm to automatically determine breakpoint placement.  Towards this goal, we have developed such an algorithm motivated by the Resultant Vector algorithm~\cite{guha2011perceptual,Talbot2011ArcLA} for  line chart aspect ratio selection, which can be found \href{
https://observablehq.com/d/b1f550d3bba19837}{as an Observable notebook}\footnote{\url{https://observablehq.com/d/b1f550d3bba19837}}.

\acknowledgments{
Support for the research is partially provided by
  NSF 1564049, 1845638, 2008295, 2106197, 452977, 1939945, 1940175; Amazon and Google research awards, and
  a Columbia SIRS award.
}

\bibliographystyle{vis/abbrv-doi}

\bibliography{references}
\end{document}